\documentclass[preprint2]{aastex}

\shorttitle{Globular clusters as modified gravity probes}
\shortauthors{X. Hernandez and M. A. Jim\'enez}

\begin{document}


\title{The outskirts of globular clusters as modified gravity probes}


\author{X. Hernandez\altaffilmark{1} and M. A. Jim\'enez}
\affil{Instituto de Astronom\'{\i}a, Universidad Nacional Aut\'{o}noma de M\'{e}xico,
Apartado Postal 70--264 C.P. 04510 M\'exico D.F. M\'exico}


\altaffiltext{1}{email:xavier@astroscu.unam.mx}


\begin{abstract}
In the context of theories of gravity modified to account for the observed dynamics of galactic systems without
the need to invoke the existence of dark matter, a prediction often appears regarding low acceleration systems:
wherever $a$ falls below $a_{0}$ one should expect a transition from the classical to the modified gravity regime.
This modified gravity regime will be characterised by equilibrium velocities which become independent of distance, 
and which scale with the fourth root of the total baryonic mass, $V^{4} \propto M$. The two above conditions are the well 
known flat rotation curves and Tully-Fisher relations of the galactic regime. Recently however, a similar phenomenology 
has been hinted at, at the outskirts of Galactic globular clusters, precisely in the region where $a<a_{0}$. Radial profiles 
of the projected velocity dispersion have been observed to stop decreasing along Keplerian expectations, and to level off at 
constant values beyond the radii where $a<a_{0}$. We have constructed gravitational equilibrium dynamical models for a number 
of globular clusters for which the above gravitational anomaly has been reported, using a modified Newtonian force law which 
yields equilibrium velocities equivalent to MOND. We  find models can be easily constructed having an inner Newtonian region 
and an outer modified gravity regime, which reproduce all observational constraints, surface brightness profiles, total masses 
and line of sight velocity dispersion profiles. Through the use of detailed single stellar population models tuned individually 
to each of the globular clusters in question, we derive estimates of the total masses for these systems. Interestingly, we find 
that the asymptotic values of the velocity dispersion profiles are consistent with scaling with the fourth root of the total 
masses, as expected under modified gravity scenarios.
\end{abstract}



\section{Introduction}

In the context of modified gravity theories, where the dynamical measurements usually ascribed to the presence of
dark matter are interpreted as evidence for a change in the form of gravity, stellar populations in the Galactic 
halo have proven to be critical test grounds. The relative proximity of many such systems makes them accessible to 
detailed photometric and spectroscopic kinematical observations, while the total absence of gas in the cases of many 
dwarf spheroidal satellites (dSph) and globular clusters, implies these objects present a relatively clean test, as
only the measured stellar component must be responsible for the measured dynamics, in the absence of any dark matter.

Galactic dwarf spheroidal satellites appear under classical gravity as the most heavily dark matter dominated systems known, 
and have been modelled extensively under MOND e.g. Sanchez-Salcedo et al. (2006), Angus (2008), Hernandez et al. (2010),
McGaugh \& Wolf (2010), Kroupa et al. (2010) for some recent examples, or equivalently, thinking of MOND as a modified
Newtonian force law, rather than a change to Newton's second law, by Mendoza et al. (2011). Sanchez-Salcedo \& Hernandez
(2007) also studied the problem of the tidal limiting of Galactic globular clusters and local dSph galaxies comparatively
under MOND and dark matter, finding both hypothesis to be consistent with the data, given the present level of observational
errors.

One generic prediction of modified gravity theories designed not to require the hypothesis of dark matter, is
that as accelerations fall below $a_{0} \approx 1.2 \times 10^{-10} m/s^{2}$, a transition should occur away 
from Newtonian gravity (e.g. MOND in Milgrom 1983, TeVeS in Bekenstein 2004, QUMOND, BIMOND in Zhao \& Famaey 2010, 
modified Newtonian force law in Mendoza et al. 2011). This modified regime will be characterised 
by equilibrium velocities for test particles orbiting within spherical mass distributions which become constant with distance, 
and which scale with mass as $V^{4} \propto M$.

In this sense, the recent studies by Scarpa et al. (2007) and Scarpa et al. (2011) appear particularly exciting, as 
they find precisely the above mentioned transition in a number of Galactic globular clusters, precisely beyond the 
radius where accelerations fall below $a_{0}$. 


The problem of Galactic globular clusters under modified gravity theories has been treated before, e.g Baumgardt et al. 
(2005) study the mean values of the velocity dispersion expected under Newtonian gravity and MOND comparatively, finding 
larger values are expected in MOND, for globular clusters in the outer halo of the Milky Way. Moffat \& Toth (2008)
however, obtain expected values for the mean velocity dispersion of GCs compatible with Newtonian expectations, and
projected velocity dispersion profiles which fall slower than Newtonian expectations, being compatible with the measurements of
Scarpa et al. (2007). Haghi et al. (2009) use a MOND N-body code to account for the effects of the external field effect of
MOND in the problem, and find that the profiles of the projected velocity dispersion of Galactic globular clusters
can serve as a test in constraining modified gravity theories. Using an analytical treatment of the problem under MOND,
ignoring the external field effect of MOND, Sollima \& Nipoti (2010) construct the MOND equivalent of analytic King profiles, 
where one limits the extent of the modelled distributions through the Galactic tides, to get models where the velocity dispersion 
profiles fall to zero. Also very recently, Haghi et al. (2011) use again a numerical MOND N-body code to show that under MOND
mean velocity dispersions of Galactic GCs, are expected to be higher than Newtonian.

In this paper we construct fully self-consistent dynamical models for Galactic globular clusters using the modified 
Newtonian force law formulation of Mendoza et al. (2011), to explore the consistency of that approach. The modelled 
density profiles are calibrated to match observed surface brightness profiles, projected velocity dispersion radial profiles, 
and total masses, inferred through a careful single stellar population modelling of the globular clusters in question, 
taking into account metalicity and age of the relevant stellar populations. We pay particular attention to reproducing
the flat outer regions of the velocity dispersion profiles, as recently measured by Drukier et al. (1998), Scarpa \& Falomo 
(2010) and Scarpa et al. (2011).

We find that fully self-consistent equilibrium models can be constructed to match all observational constraints available 
on Galactic globular clusters, including the outward flattening of the projected velocity dispersion profiles. To within 
uncertainties, the same happens under Newtonian gravity (e.g. Drukier et al. 2007, Lane et al. 2010, K\"{u}pper et al. 2010), 
where the kinematics in question are interpreted as evidence of tidal heating by the overall Galactic gravitational field, 
occurring even in the absence of prominent tidal tails, or contamination of unbound stars. It is interesting that we further 
find that the masses and asymptotic velocity dispersions of the studied globular clusters are consistent with a scaling 
$\sigma^{4} \propto M$. This last point strengthens the interpretation of the observed dynamics in the outskirts of Galactic 
GCs as evidence for modified gravity in general.

In section (2) we derive the model through which equilibrium profiles for spherically symmetric stellar populations are derived, 
under the modified Newtonian force law of Mendoza et al. (2011). In section (3) we show that such models can be easily obtained to satisfy 
all observed parameters for a sample of 8 recently observed Galactic globular clusters, all showing clearly a flattening of their 
projected velocity dispersion profiles at large radius. Our conclusions appear in section (5).

\section{Non isothermal gravitational equilibrium models}

We shall model globular clusters as populations of self gravitating stars in spherically symmetric equilibrium configurations,
(e.g. Sollima \& Nipoti 2010) under a modified Newtonian gravitational force law for test particles at a distance $r$ from the 
centres of spherically symmetric mass distributions $M(r)$:

\begin{equation}
f(x)=a_{0} x \left( \frac{1-x^{n}}{1-x^{n-1}} \right).
\end{equation}

\noindent In the above $x=l_{M}/r$, where $l_{M}=(GM(r)/a_{0})^{1/2}$. We see that when $a>>a_{0}$, $x>>1$  and $f(x)\rightarrow a_{0} x^{2}$, 
one recovers Newton's gravity force law, while when $a<<a_{0}$, $x<<1$ and $f(x)\rightarrow a_{0} x$, where an equivalent MOND 
force law $f=(G M(r) a_{0})^{1/2} /r$ is obtained. The index $n$ mediates the abruptness of the transition between the two regimes.
Notice that for a soft $n=2$ transition function one recovers the classical MOND $\mu$ function of Bekenstein (2004).

This was shown in Mendoza et al. (2011) to yield generalised isothermal gravitational equilibrium configurations
with characteristic radii, masses and velocity dispersions, $r_{g}$, $M$, $\sigma$, which smoothly evolve from the 
classical virial equilibrium of $M=\sigma^{2} r_{g} /G$ to the observed tilt in the fundamental plane of elliptical galaxies,
to the $\sigma^{4} =(M G a_{0})$ scaling of the galactic Tully-Fisher relation, as one goes from $x>>1$ to $x\sim 1$ to $x<<1$.
Also, consistency with solar system observations was found there to constrain the transition to be fairly abrupt, requiring $n>4$.
In Mendoza et al. (2011) it is also proven that a sufficient condition for Newton's theorems for spherically symmetric mass 
distributions to hold, for any modified force law, is only that $f$ can be written as a function exclusively of the variable $x$. It
is this last which we will be using in what follows, to construct equilibrium models for globular clusters. Recently, Bernal et al.
(2011) showed this modified force law model to be the low velocity limit of a formal generalisation to GR of the $f(R)$ type, providing
a theoretical basis for the model used.

The equation of hydrostatic equilibrium for a polytropic equation of state $P=K\rho^{\gamma}$ is
  
\begin{equation}
K\gamma\rho^{\gamma-2}\frac{d\rho}{dr}= -\nabla \phi.
\end{equation}

Since $\rho=(4\pi r^{2})^{-1} dM(r)/dr$, going to locally Maxwellian conditions $\gamma=1$ and $K=\sigma^{2}(r)$, 
the preceding equation can be written as:

\begin{equation}
\sigma(r)^{2} \left[  \left(\frac{dM(r)}{dr}\right)^{-1}\frac{d^{2}M(r)}{dr^{2}}-\frac{2}{r} \right]
=-a_{0} x \left( \frac{1-x^{n}}{1-x^{n-1}} \right),
\label{profile}
\end{equation}

\noindent where $\sigma(r)$ is the isotropic Maxwellian velocity dispersion for the population of stars, which is allowed to vary with 
radius, as observed in globular clusters, e.g. Sollima \& Nipoti (2010). The above is a generalisation of the treatment 
presented in Hernandez et al. (2010), which we used in the modelling of dSph galaxies, which are characterised by flat 
velocity dispersion profiles, obtaining mass models consistent with observed velocity dispersions, half mass radii and 
total masses, in the absence of dark matter. Locally Maxwellian models of this type with radially varying volumetric velocity  
dispersions can be found in e.g. Ibata et al. (2011), where conditions are further generalised to the inclusion of a varying 
radial orbital anisotropy, this last, for simplicity, we take as zero.

As an illustrative example we can take the limit $f(x)=a_{0} x^{2}$, and recover $-G M(r)/r^{2}$ for the right hand side of
equation (3), the Newtonian expression appearing for $a>> a_{0}$, or equivalently $x<< l_{M}$. If one then imposes isothermal
conditions $\sigma(r) \equiv \sigma$ and looks for a power law solution for $M(r)=M_{0}(r/r_{0})^{m}$, we get:

\begin{equation}
\sigma^{2} \left[ \frac{m-3}{r}\right]
=-\frac{ G M_{0}}{r^{2}} \left( \frac{r}{r_{0}} \right)^{m}, 
\end{equation}

\noindent and hence $m=1$, the standard isothermal halo, $M(r)=2 \sigma^{2} r/G$, having a constant centrifugal equilibrium 
velocity $v^{2}=2 \sigma^{2}$ and infinite extent. At the other limit, $a<<a_{0}$, $x>> l_{M}$, equation (3) yields:

\begin{equation}
\sigma^{2} \left[  \frac{m-3}{r}     \right]
=-\frac{[G M_{0} a_{0}]^{1/2}}{r} \left( \frac{r}{r_{0}} \right)^{m/2}.
\end{equation}

\noindent In this limit $m=0$, we obtain
$M(r)=M_{0}$ and $v^{2}= 3 \sigma^{2}=(G M_{0} a_{0})^{1/2}$, the expected Tully-Fisher scaling of the circular
equilibrium velocity with the fourth root of the mass, with rotation velocities which remain flat even after the mass
distribution has converged, rigorously isothermal halos are naturally limited in extent. It is interesting to note that 
in this limit the scaling between the circular rotation velocity and the velocity dispersion is only slightly modified 
as compared to the Newtonian case, the proportionality constant changes from 2 to 3, for the squares of the velocities.
In astrophysical units, this low acceleration limit for an isothermal halo in gravitational equilibrium yields:

\begin{equation}
\sigma =0.2 \left( \frac{M_{0}}{M_{\odot}} \right)^{1/4} km/s,
\end{equation}

\begin{equation}
v= 0.35 \left( \frac{M_{0}}{M_{\odot}} \right)^{1/4} km/s,
\end{equation}

\noindent for the velocity dispersion and the centrifugal equilibrium velocities of a halo of total mass $M_{0}$, respectively.

Also, it must be borne in mind that Galactic globular clusters are not in isolation, but orbit within the Galactic environment.
This leads to a radius beyond which the tidal forces of the Milky Way unbound the outer stars of a globular cluster, which depends
on the mass and orbital radius of the GC in question. The above tidal radius can be calculated to first order for a point mass
orbiting another point mass by equating the internal gravitational force of the GC to the derivative of the Galactic gravitational
force, we briefly recall the first order tidal limit calculation of: 

\begin{equation}
 \left. \frac{d F_{ext}(R)}{dR} \right|_{R_{0}}  \Delta r = F_{int}, 
\end{equation}

\noindent which leads to the tidal density stability condition of $\rho_{s}>\overline{\rho}$ for the density of a satellite 
of extent $\Delta r$ and mass $M_{s}$ orbiting at a distance $R_{0}$ from the centre of a spherical mass distribution $M(R)$ 
having an average matter density $\overline{\rho}$ internal to $R_{0}$ resulting in a gravitational force $F_{ext}(R)$, where 
the internal gravitational force under Newton is $F_{int}= -GM_{s}^{2}/ (\Delta r)^{2}$, under the assumption $\Delta r <<R$. 
The equivalent calculation under the force law given by the $a<<a_{0}$ limit of $F=-(G M a_{0})^{1/2}/R$ is given by:

\begin{equation}
\frac{(G M(R) a_{0})^{1/2}}{R^{2}}\Delta r  = \frac{(G M_{s} a_{0})^{1/2}}{\Delta r},
\end{equation}

\noindent leading to:

\begin{equation}
\rho_{s}> \left(\frac{\Delta r} {R} \right)  \overline{\rho},
\end{equation}

\noindent as the equivalent of the classical tidal density criterion. In terms only of masses and radii we obtain the tidal radius as:

\begin{equation}
r_{t}=R \left( \frac{M_{s}}{M(R)} \right)^{1/4}
\end{equation}

Since the spatial extent of globular clusters will always be much smaller that their Galactocentric radii, equation(10)
shows that under the modified gravitational force law explored here, GCs will be much more robust to tides than 
under Newtonian gravity. As an example, for a $3 \times 10^{5} M_{\odot}$ GC orbiting $10 kpc$ from a $1.6 \times 10^{11} M_{\odot}$
mass, corresponding under eq.(7) to the $220 km/s$ of the Galactic rotation curve, the tidal radius comes to $370 pc$. 
This is relevant as it shows that treating GCs as isolated systems is a self-consistent assumption, within the approach of 
the modified force law being considered. Recent examples of dynamical modelling of Galactic GCs treated as isolated
systems can be found in e.g. Sollima \& Nipoti (2010) and Haghi et al. (2011).

Taking initial conditions $M(r)\rightarrow 0$, $dM(r)/dr \rightarrow 4\pi r^{2}\rho_{0}$, when $r\rightarrow 0$, a constant central 
density $\rho_{0}$, we can solve the full second-order differential equation (\ref{profile}) for $M(r)$ through a numerical finite 
differences method, once a model for $\sigma(r)$ is adopted.

Solving equation (\ref{profile}) yields the volumetric profiles for the density and mass, $\rho(r)$, $M(r)$. $\rho(r)$
is then projected along one dimension to obtain a projected surface density mass profile, $\Sigma(R)$. In all that follows we shall
use $r$ for radial distances in 3D, and $R$ for a projected radial coordinate on the plane of the sky. $\Sigma(R)$ can be
compared to observed surface brightness profiles once a mass-to-light ratio is assumed. Given the appearance of mass 
segregation processes in the dynamically evolved stellar populations of Galactic globular clusters, volumetrically,
the mass-to-light ratio of a real GC will never be strictly constant. However, given the ages of the stellar populations
involved, only a narrow range of stellar masses remain, and further, observed surface brightness profiles are projections 
on the plane of the sky of the volumetric density profiles. Thus, observed surface brightness measurements towards the 
central regions of GCs imply an integration across the entire foreground and background of any observed cluster. This in
turn means that the regions where mass segregation is strongest, the central ones, contribute to the projected surface 
brightness profiles only after being averaged out over lines of sight traversing entirely the cluster in question. 
Even after this averaging process, the high densities found towards the centre still surely imply M/L values which 
strictly must be a function of radius, even for the surface brightness profiles $I(R)$. Still, for the reasons described
above, it is frequent to compare modelled $\Sigma(R)$ mass surface density profiles directly to observed projected
surface brightness profiles through the use of constant M/L values, e.g. Jordi et al. (2009), Haghi et al. (2011). In what follows
we will also perform such comparisons assuming constant M/L values.

\begin{figure}[!t]
\plotone{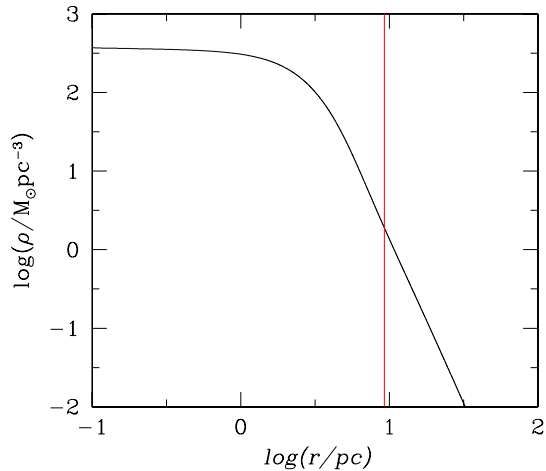}
\caption{The figure shows the volumetric density profile for a sample model. The vertical line gives the point where $x=1$. 
Notice that the asymptotic $\rho(r)$ profile at large radii is steeper than the $r^{-2}$ of the Newtonian case, resulting 
in a finite total radius and total mass for the configuration.}
\end{figure}

\begin{figure}[!t]
\plotone{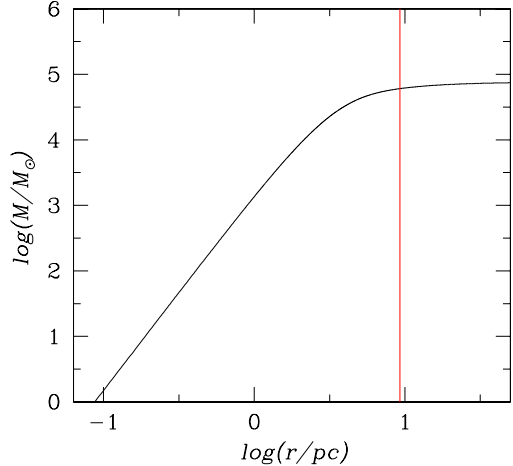}
\caption{The figure shows the volumetric mass profile for the same sample model of fig.(1). The vertical line gives the point
where $x=1$. The convergence to a total mass is evident, despite not having introduced any externally determined tidal radius.}
\end{figure}

Finally, a projected velocity dispersion  
profile $\sigma_{p}(R)$ is constructed through a volume density weighted projection of the volumetric velocity dispersion 
profile of a model, $\sigma(r)$. Therefore, a proposed model for $\sigma(r)$ does not directly give a projected $\sigma_{p}(R)$
profile, which the model only yields after having solved for the detailed density structure, and the subsequent mass weighted
projection of the proposed  $\sigma(r)$.

Observations provide only the density weighted projected profiles of $\sigma(r)$, integrated along the line of sight, $\sigma_{p}(R)$, not the 
volumetric $\sigma(r)$ profiles which we require in equation (\ref{profile}). We shall therefore adopt a parametric form for the 
volumetric velocity dispersion profile, and adjust the parameters to obtain a match to the observed globular cluster properties. 
For this we use:

\begin{equation}
\sigma(r)=\sigma_{1} \mathrm{exp}\left(-\frac{r^{2}}{r_{\sigma}^{2}}\right)+\sigma_{0}.
\label{sigma}
\end{equation}

In the above $\sigma_{0}$ is given directly by the observations as the asymptotic value of the measured projected
velocity dispersion profile for each globular cluster, as at large $R$ radii, projection effects tend to zero and
$\sigma_{p}(R) \rightarrow \sigma(r)$. This leaves us with three model parameters to determine: $\rho_{0}$,
$\sigma_{1}$ and $r_{\sigma}$, which we fit to match the observed projected velocity dispersion profiles, as well as central 
projected velocity dispersion values for each observed globular cluster treated, and comparing the resulting model 
$\Sigma(R)$ profiles to the observed surface brightness profiles of any given GC, under the requirement that the
$M/L$ ratios used be consistent with detailed inferences from the stellar evolutionary and IMF studies of
McLaughlin \& van der Marel (2005).

We end this section presenting in figures (1) and (2), a sample model. In figure (1) we show the volumetric density
profile, which is qualitatively similar to a cored isothermal profile, with the difference that the asymptotic 
$\rho(r)$ profile at large radii is steeper than the $r^{-2}$ of the Newtonian case. This results in finite total masses and 
finite half-mass radii even for the asymptotically flat $\sigma(r)$ volumetric profiles we adopt, in contrast with the situation in 
classical gravity, where infinitely extended mass profiles would appear. This finite profiles are also what appears under the type of 
modified gravity laws we are treating, even for rigorously isothermal $\sigma(r)=\sigma_{0}$ equilibrium configurations, as already pointed 
out in Hernandez et al. (2010) and Mendoza et al. (2011) and as shown in the developments following eq. (5).

Figure (2) gives the corresponding volumetric radial mass profile for the same sample model, indicating again with a vertical line 
the threshold where $x=1$. We see that only $20 \%$ of the total model mass lies beyond $x=1$. This fraction is distributed over a 
much larger area having much smaller projected mass surface densities than the central regions interior to the $x=1$ threshold, 
the ones which are much more easily observed, and over which Newtonian gravity accurately holds. 

These two figures illustrate the 
physical consistency of the model, a positive isothermal volumetric velocity dispersion is assumed, integration of eq.(3) then yields 
a volumetric density profile essentially consistent with what would appear under Newtonian dynamics in the region interior to $x=1$, 
where the force law converges precisely to the standard expression. At large radii however, the density profile increasingly steepens 
and naturally reaches $\rho(r)=0$ at a well defined total radius, as is evident from the convergence seen in figure (2). Thus, the 
distribution function is necessarily positive throughout the modelled structure, and goes to zero at a well defined outer radius. 
In the following section we give best fit models for 8 recently observed Galactic globular clusters.

\section{Modelling observed globular clusters}

\begin{figure*}[!t]
\includegraphics[width=17cm,height=18.3cm]{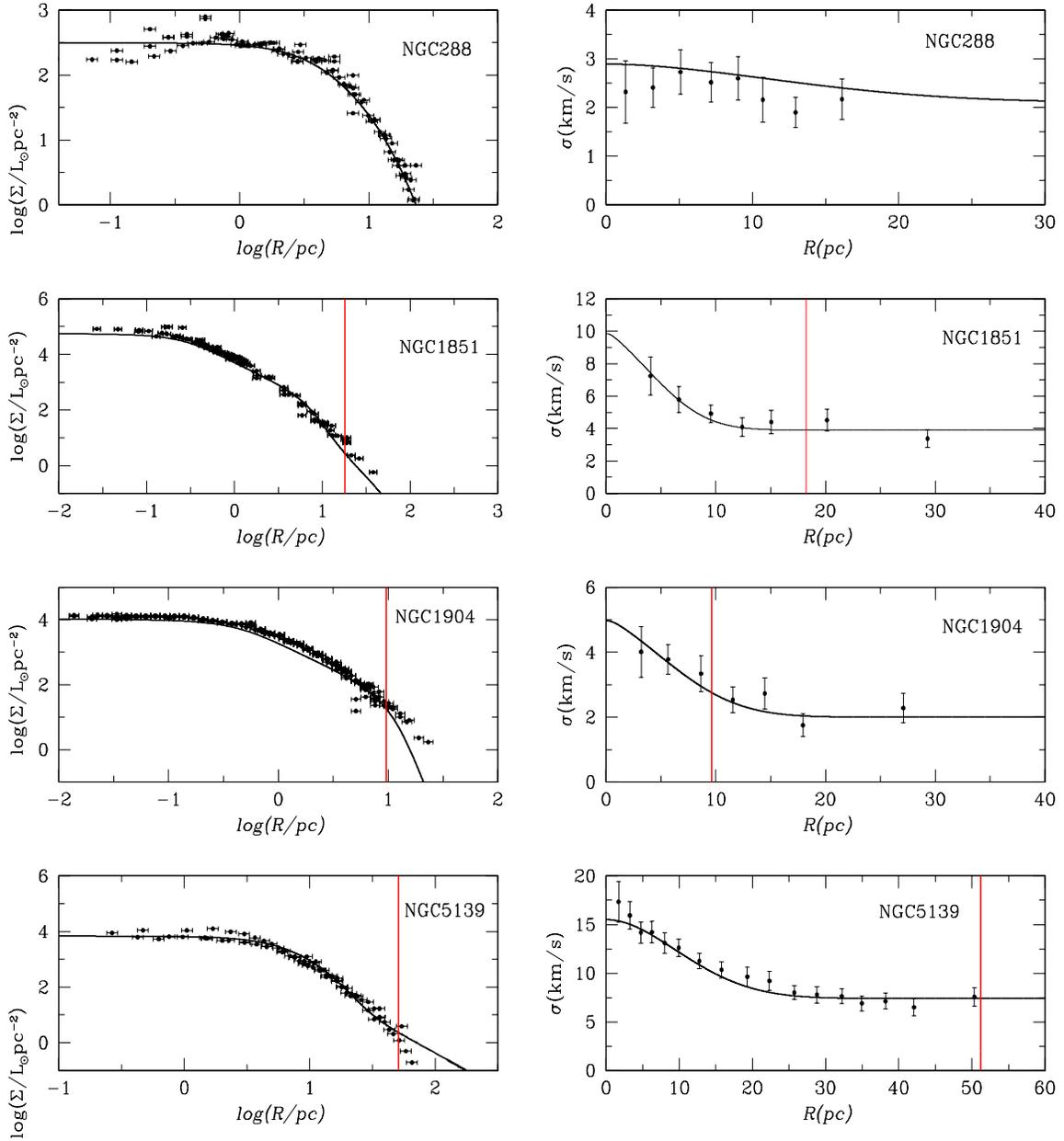}
\caption{The figure compares the resulting model projected surface brightness profiles and resulting model projected
velocity dispersion profiles to the corresponding observed quantities for the first four globular clusters studied. In all cases
the assumed $M/L$ values where within the ranges given by McLaughlin \& van der Marel (2005) for each individual GC through
detailed stellar population modelling. The vertical line gives the point where $x=1$ and the modified force law used
shifts from the classical Newtonian form to the MONDian character of eq.(5).}
\end{figure*}

\begin{figure*}[!t]
\includegraphics[width=17cm,height=18.3cm]{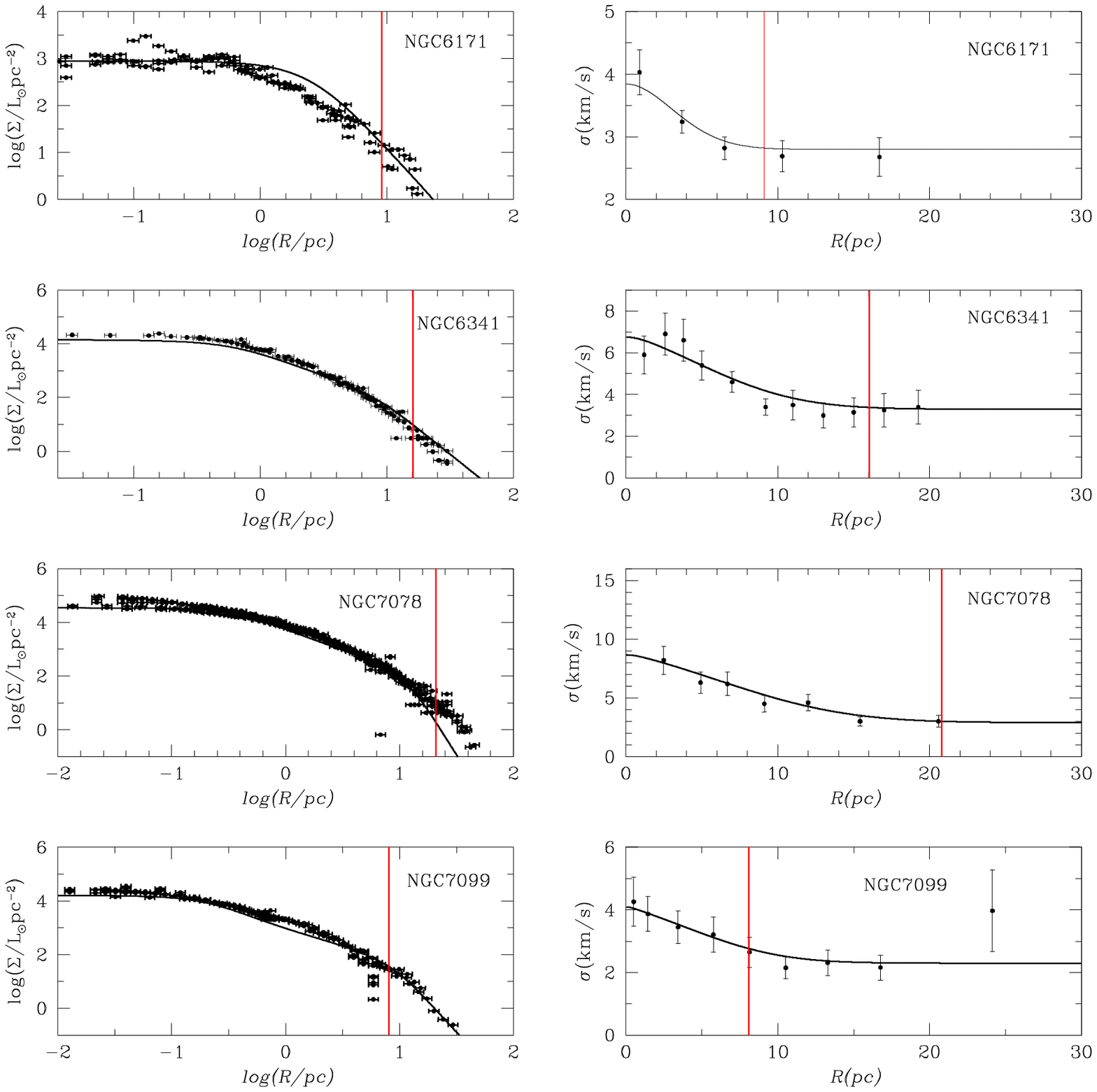}
\caption{The figure compares the resulting model projected surface brightness profiles and resulting model projected
velocity dispersion profiles to the corresponding observed quantities for the last four globular clusters studied. In all cases
the assumed $M/L$ values where within the ranges given by McLaughlin \& van der Marel (2005) for each individual GC through
detailed stellar population modelling. The vertical line gives the point where $x=1$ and the modified force law used
shifts from the classical Newtonian form to the MONDian character of eq.(5).}
\end{figure*}


We now present best fit models constructed as described above, optimised to match observed projected surface brightness profiles,
under $M/L$  values consistent with the ranges given by McLaughlin \& van der Marel (2005) for each individual GC modelled,
and projected velocity dispersion measurements for a series of Galactic globular clusters. The data for the projected velocity
dispersion profiles we take from Drukier et al. (1998), Scarpa et al. (2004), Scarpa et al. (2007a),  Scarpa et al. (2007b),
Scarpa \& Falomo (2010) and Scarpa et al. (2011), and the surface brightness profiles from Trager et al. (1995). Total masses we infer from the 
integration of the observed surface brightness profiles, assuming again  $M/L$  values consistent with the ranges given by 
McLaughlin \& van der Marel (2005) for each individual GC modelled, detailed single stellar population models tuned to the 
inferred ages and metallicities of each of the clusters we model, and checked for consistency through the synthetic HR diagram 
construction procedure described in Hernandez \& Valls-Gabaud (2008). 


For this last, the relevant ages and metalicities we take from the literature: Salaris \& Weiss (2002); Kraft \& Ivans (2003) for 
NGC 288 and NGC 6341, Salaris \& Weiss (2002); Rutledge et al. (1997) for NGC 6171, Salaris \& Weiss (2002) and McNamara et al. (2004); 
Kraft \& Ivans (2003) for NGC 7078 and Samus et al. (1995); Harris (1996) for NGC 7099, respectively. In obtaining inferred total masses, 
confidence intervals were assigned by considering the full range of $M/L$ values given by McLaughlin \& van der Marel (2005) for each 
individual GC, for the full range of plausible stellar models considered in that study, for each of the GCs we treat. The final results 
for the mass to light ratios appear in the table at the end of this section.

The final resulting density weighted projected velocity dispersion profiles, $\sigma_{p}(R)$, together with the resulting surface brightness
profiles for the 8 GCs treated appear in figures (3) and (4), where the vertical line indicates the point where $a=a_{0}$. The models accurately fit 
the observed projected surface brightness profiles of the clusters, the central value of the measured projected velocity dispersion, as 
well as the observed profiles for $\sigma_{p}(R)$. 

We see five cases, NGC 288, NGC 1851, NGC 5139,  NGC 6341 and NGC 7099 where the models very accurately 
reproduce all light distribution and velocity dispersion profiles, these last in fact agree with the observations in all cases, to within 
reported uncertainties. For NGC 1904 and NGC 6171 a very slight mismatch appears between the models and the observations of the surface brightness 
profiles at values of $R\sim$ a few pc, and for NGC 7078, over the outer regions, precisely beyond $x=1$, the model falls below the observed 
surface brightness profile. These two last points could be evidence of the failure of some of the assumptions used in these cases, probably
the presence of velocity anisotropy in the dynamics of the stars in question. Alternatively, a slight tuning of the assumed 
volumetric $M/L$ ratios with radius, within entirely plausible ranges, would improve the fits. Rather than introduce a further degree
of freedom, we prefer to show that highly adequate fits are easily obtained, for the simplest $M/L=$ constant assumption.

We see also the steepening of the surface density brightness profile 
towards the edge of the clusters, particularly for NGC 288 and NGC 7099. Under Newtonian gravity this feature would be interpreted as a tidal 
radius, while under the modified force law scheme treated here, it is a natural consequence of the change in gravitational regime, leading
to finite matter distributions when $a<a_{0}$ and $\sigma(r)$ tends to a constant cf., eq.(5).


A model with a degree of radial 
orbital anisotropy which varies with radius is an entirely plausible alternative, implying the introduction of a further free function
which allows to more accurately reproduce the surface brightness profiles. We have also chosen not to complicate the model with the inclusion
of a $\beta(r)$ profile, and preferred to show that under the simplest isotropic construction, the modified force law we test
is capable of adequately reproducing all the observations in five of the cases studied, although slight mismatches appear in three 
other cases. In particular, for NGC 7078, our model requires more mass at large radii, but without an increase in $\sigma(r)$ in that
region. This can be trivially accommodated by an increase of kinetic energy at large radius which should not be evident in the
observed velocity dispersion, through the inclusion of radial motion, a degree of orbital radial anisotropy appearing beyond
20 pc.

\begin{figure}[!t]
\plotone{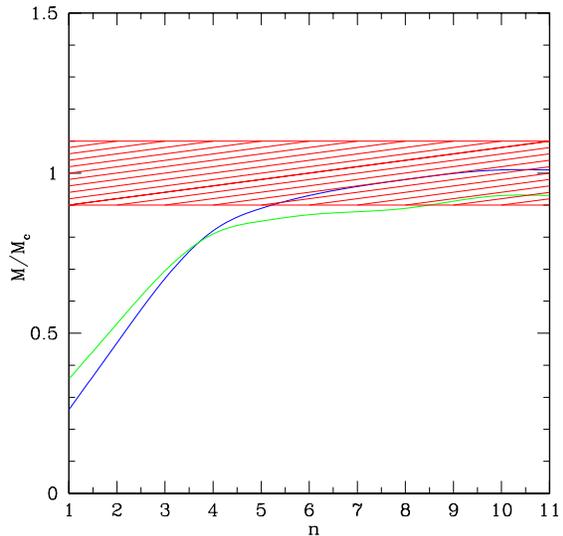}
\caption{Dependence of resulting model mass on the value of the force law index $n$. The ratio of the model mass to the
inferred globular cluster mass is shown for two representative clusters, with the shaded region giving the range allowed
by direct stellar population studies of the clusters shown, NGC 288 and NGC 6341. This illustrates the need for a sharp 
transition in  the force law used.}
\end{figure}

Obtaining a good match required taking a high exponent of $n=10$ in equation (3), i.e., a relatively sharp transition 
for the generalised force law, which however, remains continuous and differentiable at all points, by construction.
This is in agreement with observed upper limits to deviations from Newtonian dynamics
at the solar system, models for the Galactic rotation curve, and equilibrium models for the local dSphs, as shown in 
Mendoza et al. (2011) to apply for the modified force law tested here, provided $n>4$. Essentially this implies that the transition 
between the Newtonian and the ``MONDian'' regimes must be fairly abrupt. In fact, given the form of equation (3), the final model becomes 
independent of $n$ for any value of $n$ higher than 10.

We now illustrate the dependence of the models obtained to the assumed force law index $n$. Figure (5) shows the ratio between
the resulting model mass, $M$, to the observed cluster mass, $M_{c}$, as a function of the value of $n$ used, for best fit models 
for two illustrative cases, NGC 288 and NGC 6341, upper and lower curves at the right end, respectively. The shaded region 
gives the allowed range of globular cluster mass for a particular choice of stellar population parameters in McLaughlin \& van 
der Marel (2005), if the systematic uncertainties were included, the shaded region would extend, but only upwards, by a factor of
about 2. We see that obtaining a total model mass in accordance with direct stellar population studies requires taking a value of
$n>5$ for NGC 288, and $n>8.5$ for NGC 6341. The particular minimum value of $n$ required for the other clusters in the sample
varies somewhat, although most require values $n>8-9$ to reach the minimum inferred masses. As this parameter can not be expected
to change from cluster to cluster, we have taken $n=10$ in all cases. 

Notice that this value results in a equivalent MOND
$\mu$ transition function which is fairly abrupt. For comparison, most MOND transition functions proposed imply a softer
transition from the Newtonian to the MONDian regimes, e.g. the $\mu$ function of Bekenstein (2004) correspond exactly to 
the force law treated here, with $n=2$. A similar result has recently been presented by Qasem (2011), who analyses laser
lunar ranging data to constrain the allowed departures from Newtonian gravity at the scale studied, and finds the data
rule out most proposed MOND transition functions, while our proposal of Mendoza et al. (2011) remains consistent with the 
test performed. This completely independent study supports the constraints of figure (5), implying that the transition from 
the Newtonian regime to the MONDian one is probably steeper than commonly thought.

\begin{figure}[!t]
\plotone{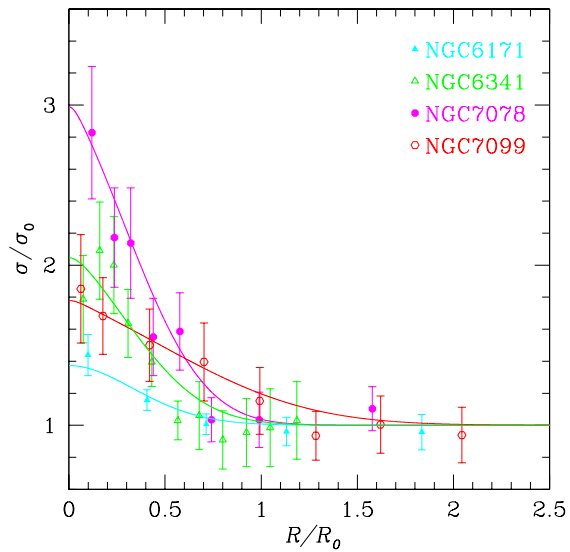}
\caption{The curves represent the projected velocity dispersion profiles normalised to $\sigma_0$ and to $R_{0}$ of our models for 
four representative examples.}
\end{figure}

Figure (6) now gives the radial profiles of $\sigma_{p}(R)$ for 4 of the Galactic globular clusters in the sample, where we 
have normalised the velocity dispersions to their asymptotic values, and the radial coordinate to the values where the 
condition $x=1$ ($a=a_{0}$) is met in each of the models, $R_{0}$. Measurements with confidence intervals appear represented 
by different symbols for each GC. We see again that $\sigma_{p}(R)$ profiles consistent 
with observations are easily obtained, while fitting simultaneously surface brightness profiles, to within 
the error bars, where we have also assumed a $20 \%$ uncertainty in all distance determinations. The four clusters
included in this figure illustrate the range of behaviours found, from NGC 7078 which shows a very large variation in $\sigma_{p}(R)$ 
from the centre to the outskirts, implying a large Newtonian inner region and only a slight MONDian outer zone, to clusters with only 
small variations in $\sigma_{p}(R)$ like NGC 6171 and NGC 7099. An extreme case is NGC 288 with a $\sigma_{p}(R)$ profile which is also 
consistent with observations, but which can not be plotted in figure (6), as that low density cluster lies in the $a<a_{0}$ regime 
throughout, and therefore $R_{0}$ is not defined for it. 

Notice that no external field effect of MOND was included in the dynamical modelling performed, as was also not included 
in e.g. Sollima \& Nipoti (2010). That accurate mass and velocity dispersion models can be thus constructed, is suggestive
of a modified gravity formulation where no such effect appears.

Finally, in figure (7) we plot inferred values of the total masses and asymptotic values of projected velocity dispersions
for the 8 globular clusters we study, with their corresponding confidence intervals. The solid line shows the best fit 
power law dependence for this plot, excluding Omega Cen. from the fit. Notice that the quantities plotted here are completely 
independent of any dynamical modelling, being merely measured velocity dispersion values in the outskirts of the clusters in 
question, and total masses as inferred from total luminosities and models of their observed stellar populations.

Despite the small ranges in parameters covered by the clusters studied, it is clear that the data are consistent with the 
generic modified gravity prediction of $\sigma_{0} \propto M^{1/4}$, the slope of the solid line being statistically determined as 
$0.31 \pm 0.06$. For comparison, the dashed line gives the best slope 1/4 fit, as the numbers above show, a consistent
description of the data. Omega Cen. was excluded from the fit, as it is usually acknowledged as an outlier. The complex stellar populations
it presents, its abnormally high mass, together with the possible presence of an intermediary mass black hole in its
centre probably indicate a complex dynamical formation history. Still, its inclusion in the power law fit of figure (7)
only slightly modifies the resulting slope to $0.32 \pm 0.04$.

\begin{figure}[!t]
\plotone{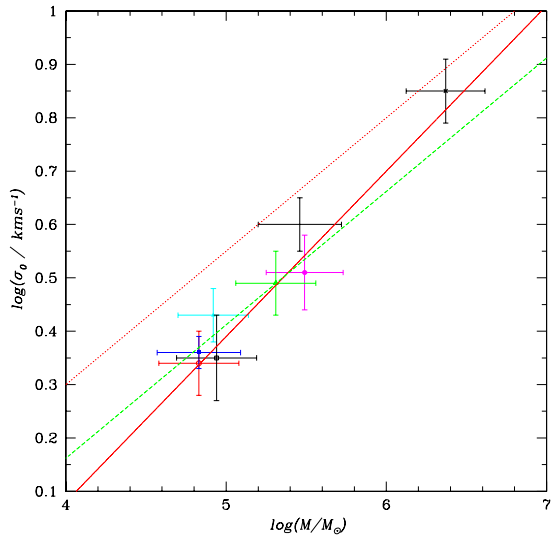}
\caption{The points with error bars represent the values of $\sigma_0$ and total stellar mass for the 8 globular clusters we study. 
The solid line gives the best fit power law, having a slope of $0.31 \pm 0.06$, excluding Omega Cen. from the fit. If Omega Cen
is included in the fit, the slope comes to $0.32 \pm 0.04$. The dotted curve shows the prediction of eq.(6) for isothermal 
systems fully in the $a<<a_{0}$ regime, while the dashed line gives the best slope 1/4 fit.}
\end{figure}

Figure (7) lends credibility to the interpretation of the outer flattening of the observed velocity dispersion profiles as evidence 
for modifications in the law of gravity. Under classical gravity, one needs to invoke further causal correlations to account for 
the clear trend appearing in figure (7), a more contrived scenario than simply the natural consequence of a shift towards MONDian 
dynamics in globular clusters at scales where $a<a_{0}$. As a comparison, the dotted line in figure (7) shows the prediction 
of eq.(6) for isothermal systems fully in the $a<< a_{0}$ regime.

\begin{table*}
  \caption{Model parameters for the globular clusters treated.}
  \begin{tabular}{@{}llllllll@{}}
  \hline
  \hline
   Globular Cluster & $\,\,\,log(\rho_{0})\,\,$ & $\sigma_{1}\, $ & 
$\sigma_{0}$  & \,\,$r_{\sigma}\,$ & $(M/L)_{M}$ & $(M/L)_{GC}$\\
 \hline
 & & & & & & &\\
 NGC 288      &\,\, $2.15      $   & 0.83  & 2.1    & 18.0  & \,\,\,\,\,2.0  & $3.03 \pm 1.12$\\
 & & & & & & &\\
 NGC 1851     &\,\, $5.65      $   & 6.00  & 3.9    & 7.0   & \,\,\,\,\,2.0  & $3.00 \pm 1.19$\\
 & & & & & & &\\
 NGC 1904     &\,\, $4.70      $   & 3.00  & 2.0    & 9.0   & \,\,\,\,\,1.73  & $2.73 \pm 1.00$\\
 & & & & & & &\\
 NGC 5139     &\,\, $3.60      $   & 8.90  & 7.4    & 15.0   & \,\,\,\,\,2.0  & $2.68 \pm 0.98$\\
 & & & & & & &\\
 NGC 6171     &\,\, $3.52      $   & 1.20  & 2.8    & 5.0   & \,\,\,\,\,2.5  & $3.20 \pm 1.20$\\
 & & & & & & &\\
 NGC 6341     &\,\, $4.68      $   & 3.50  & 3.3    & 9.0   & \,\,\,\,\,1.63 & $2.55 \pm 0.95$\\
 & & & & & & &\\
 NGC 7078     &\,\, $4.98      $   & 5.80  & 2.9    &11.0   & \,\,\,\,\,1.7  & $2.51 \pm 0.90$\\
  & & & & & & &\\
 NGC 7099     &\,\, $4.92      $   & 1.80  & 2.3    & 8.0   & \,\,\,\,\,1.65 & $2.60 \pm 0.90$\\
 
\hline

\end{tabular}
\begin{flushleft} 
$\rho_{0}$ gives the central values of the stellar density used in each model in units of 
$M_{\odot} pc^{-3}$, while $\sigma_{0}$, $\sigma_{1}$ and $r_{\sigma}$ give the parameters of the volumetric 
velocity dispersion profile used, in units of $km/s$ and $pc$, respectively. $(M/L)_{M}$ gives the mass-to-luminosity
ratios used in each model, and $(M/L)_{GC}$ gives the corresponding values reported by McLaughlin \& van der Marel (2005) 
for each cluster through single stellar population modelling using age and metallicity parameters as appropriate for each.
\end{flushleft}
\end{table*}

That the GCs treated lye somewhat below the prediction of eq.(6) is not surprising, as they are by no means structures 
fully in the $a<<a_{0}$ regime, most of the mass is partly in the Newtonian regime, or in the transition between both,
hence the asymptotic velocity dispersions slightly off the results of eq.(6).

We acknowledge that a number of the assumptions going into the modelling remain as such, it is entirely plausible that 
some might not strictly apply. As already discussed, the case of NGC7078 could be indicative of a certain degree of
orbital radial anisotropy appearing towards the outskirts of some of these systems. Also, the projected $M/L$ values might 
have some radial variations, or the flattening of the projected velocity dispersion profiles might simply reflect the presence 
of contaminating unbound stars, or the effects of tidal heating, as assumed under Newtonian interpretations of GC structure 
e.g. Drukier et al. (2007) or Lane et al. (2010). Under Newtonian dynamics, perfectly self-consistent models can be constructed, 
often leading to even better fits than those resulting from modified gravity approaches, e.g. Ibata et al. (2011). It is the 
new results of figure (7), which are independent of any gravitational modelling, which we find most suggestive of the 
modified gravity interpretation, as under the standard approach, further mechanisms must be invoked to explain the appearance 
of a "Tully-Fisher" relation for GCs. 

We conclude this section with table (1), which summarises the observed parameters used and the model parameters fitted
for each of the 8 galactic globular clusters in this paper. We note also that the models are relatively insensitive to 
small changes in the values of $\rho_{0}$ used, only changes of factors $\sim 3$ and above in this model parameter result
in significant changes.

\section{Conclusions}

We show that for Galactic globular clusters, spherically symmetric equilibrium models can be constructed using a modified 
Newtonian force law which reproduces the MOND phenomenology, which naturally satisfy all observational constraints available.

The resulting models are typically characterised by a Newtonian inner region, which smoothly transits to a modified gravity
outer region on approaching the $a=a_{0}$ threshold. The resulting internal velocity dispersion profiles correspondingly 
transit from an inner radially decaying region to an outer flat velocity dispersion one.

By comparing against careful single stellar population modelling of the globular clusters studied to derive total stellar 
mass estimates, we show that the asymptotic values of the measured velocity dispersion profiles, $\sigma_{p}(R\rightarrow \infty)$, 
and total masses for these systems, $M$, are consistent with the generic modified gravity prediction for a scaling 
$\sigma_{p}^{4}(R\rightarrow \infty) \propto M$.

\acknowledgments

The authors acknowledge the constructive criticism of an anonymous referee as valuable towards reaching a clearer 
and more complete manuscript. Xavier Hernandez acknowledges financial assistance from UNAM DGAPA grant IN103011-3. 
Alejandra Jimenez acknowledges financial support from a CONACYT scholarship.

\end{document}